\documentclass[fleqn,usenatbib]{mnras}

\usepackage{newtxtext,newtxmath}
\usepackage[dvipsnames]{xcolor}
\usepackage{float}
\usepackage[utf8]{inputenc}
\usepackage{graphicx}

\usepackage{soul}
\newcommand{\bomu}{\ensuremath{\boldsymbol{\mu}}}

\newcommand{\boX}{\ensuremath{\boldsymbol{x}}}
\newcommand{\boC}{{\sf{C}}}

\newcommand{\mathd}{\mathrm{d}}

\newcommand{\both}{\ensuremath{\boldsymbol{\theta}}}
\newcommand{\obsboX}{\ensuremath{\boX_{\mathrm{o}}}}

\newcommand{\sfC}{{\ensuremath{\sf{C}}}}

\newcommand{\fatx}{\ensuremath{\boldsymbol{x}}}

\newcommand{\calP}{\ensuremath{\mathcal{P}}}

\newcommand{\calL}{\ensuremath{\mathcal{L}}}

\newcommand{\calG}{\ensuremath{\mathcal{G}}}

\usepackage[T1]{fontenc}

\DeclareRobustCommand{\VAN}[3]{#2}
\let\VANthebibliography\thebibliography
\def\thebibliography{\DeclareRobustCommand{\VAN}[3]{##3}\VANthebibliography}

\usepackage{graphicx}	% Including figure files
\usepackage{amsmath}	% Advanced maths commands

\title[Baryonic feedback solution for non-Gaussianity]{Stage IV baryonic feedback correction for non-Gaussianity inference}

\author[Grand\'on \& Sellentin]{Daniela Grand\'on,$^{1}$\thanks{d.i.grandon.silva@math.leidenuniv.nl}
Elena Sellentin,$^{1,2}$
\\
$^{1}$Mathematical Institute, Leiden University, Snellius Gebouw, Niels Bohrweg 1, NL-2333 CA Leiden, The Netherlands\\
$^{2}$Leiden Observatory, Leiden University, Oort Gebouw, Niels Bohrweg 2, NL-2333 CA Leiden, The Netherlands\\
}
\date{Accepted XXX. Received YYY; in original form ZZZ}

\pubyear{2023}

\begin{document}
\label{firstpage}
\pagerange{\pageref{firstpage}--\pageref{lastpage}}
\maketitle

\begin{abstract}
Non-Gaussian statistics of the projected weak lensing field are powerful estimators that can outperform the constraining power of the two-point functions in inferring cosmological parameters. This is because these estimators extract the non-Gaussian information contained in the small scales. However, fully leveraging the statistical precision of such estimators is hampered by theoretical uncertainties, such as  those arising from baryonic physics. Moreover, as non-Gaussian estimators mix different scales, there exists no natural cut-off scale below which baryonic feedback can be completely removed.
We therefore present a Bayesian solution for accounting for baryonic feedback uncertainty in weak lensing non-Gaussianity inference. Our solution implements Bayesian model averaging (BMA), a statistical framework that accounts for model uncertainty and combines the strengths of different models to produce more robust and reliable parameter inferences. We demonstrate the effectiveness of this approach in a Stage IV convergence peak counts analysis, including three baryonic feedback models. We find that the resulting BMA posterior distribution safeguards parameter inference against biases due to baryonic feedback, and therefore provides a robust framework for obtaining accurate cosmological constraints at Stage IV precision under model uncertainty scenarios.
\end{abstract}


%%%%%%%%%%%%%%%%%%%%%%%%%%%%%%%%%%%%%%%%%%%%%%%%%%

%%%%%%%%%%%%%%%%% BODY OF PAPER %%%%%%%%%%%%%%%%%%
\section{Introduction} 
\label{sec:intro}
The formation of cosmic structures is determined by gravity and the expansion history of the Universe. In the late Universe, structure growth has evolved into the non-linear regime, resulting in matter being distributed as a non-Gaussian random field. Weak gravitational lensing is particularly affected by such non-linearities, as its effects are driven by the total matter distribution. Consequently, estimators that capture the non-Gaussian features in the lensing field are valuable tools for extracting additional cosmological information contained in the small (non-linear) scales. 

In recent years, Stage III weak lensing surveys such as the Kilo Degree Survey\footnote{kids.strw.leidenuniv.nl} (KiDS, \cite{KiDS:2015vca, KiDS:2020suj}), the Dark Energy Survey\footnote{www.darkenergysurvey.org} (DES, \cite{DES:2016jjg, DES:2017, DES:2021bvc, DES:2021vln}), and the Hyper Suprime Cam\footnote{www.naoj.org/Projects/HSC} (HSC, \cite{HSC2017, HSC}) have implemented inference from non-Gaussian estimators. These analyses show that such estimators can tighten cosmological constraints compared to inference of the power spectrum alone (e.g. \cite{martinet2018kids,shan2018kids,gatti_chang,Martinet:2020omp,zurcher2022dark,liu2023cosmological,Gabriela2023,Thiele:2023gqr,  Cheng:2024kjv, Harnois-Deraps2024,Grandon:2024pek}). Amongst the most studied non-Gaussian statistics we find Minkowski functionals \citep{PhysRevD.85.103513,Marques:2018ctl,parroni2020, Grewal:2022qyf}, peak counts \citep{ Liu:2014fzc, Kacprzak2016,Li:2018owg,Ajani:2020dvu, 2021MNRAS.506.1623H, 2023A&A...671A..17A, Davies:2024nlc}, minimum counts \citep{Coulton:2019enn, Gabriela2023}, the one-point probability density function \citep{ Liu:2018pdf,2020MNRAS.492.3420B,Thiele:2020pdf,2021MNRAS.505.2886B, Giblin:2022ucn,Barthelemy:2023mer, 2024arXiv240509651C}, scattering transform coefficients \citealp{Cheng_2020, Cheng_2021,Valogiannis:2021chp}, and starlet $\ell_1$ norm \citep{2021A&A...645L..11A, 2023A&A...672L..10A}. 
Some of the non-Gaussian estimators have also been studied at Stage IV precision, for mock data of \textit{Euclid} \citep{Euclid} and Vera Rubin Observatory Legacy Survey of Space
and Time \citep{LSST:2008ijt}. These cosmological forecasts predict that a joint analysis of non-Gaussian statistics and two-point functions can improve the constraints on cosmological parameters by a factor of 2 to 3 compared to using the two-point function alone \citep{2023A&A...675A.120E}. However, a full Bayesian parameter inference analysis is needed to study cosmological constraints from noisy real data and the parameter biases that can arise due to unmodeled systematic effects. 

In this context, baryonic feedback is one of the most important astrophysical systematic in weak lensing analysis. It describes how cosmic matter fields are subject not only to gravitational collapse but also to matter redistribution by stellar and galactic processes from intermediate to small scales. These processes include supernova feedback, star formation, gas cooling, and active galactic nuclei (AGN) feedback, amongst others. Given the complexity in the modeling of baryonic physics, many hydrodynamic simulations suites are required. They differ in many aspects, including specific calibration strategies of sub-grid parameters, box size and resolution \footnote{Hydrodynamic simulations also differ in initial conditions, hydrodynamic solvers, and number of sub-grid parameters.}  \citep{2010MNRAS.402.1536S,2011MNRAS.415.3649V, 2014Natur.509..177V, 2014MNRAS.444.1453D, 2015MNRAS.450.1349K,2016MNRAS.461L..11H,2018MNRAS.475..676S, McCarthy:2016mry, 2017MNRAS.472.2153P,Mccarthy:2017yqf,2023MNRAS.526.5494M}. 

Results from \cite{Chisari:2019tus} and \cite{vanDaalen:2019pst} reveal a significant discrepancy in the amplitude of the effects of baryonic feedback across hydrodynamic simulations (and the scales at which these effects become important). This is because baryonic feedback encapsulates a list of complicated processes and to date no consensus on a single outstanding model has been reached. Accounting for a multitude of baryonic feedback models in a non-Gaussianity analysis is hence paramount, if the inferred primary cosmological parameters are to be unbiased.
In \cite{Grandon:2024pek} we show the impact of baryons on several non-Gaussian estimators based on Subaru Hyper Suprime-Cam Y1 mock data. We demonstrate that unmodelled baryonic physics lead to $\sim 1\sigma$ biases on the structure growth parameter $S_8$ when including the smallest scales. However, \cite{sembolini2013}, \cite{Coulton:2019enn} and \cite{Martinet:2020omp}  show that this effect is forecasted to be severe at Stage IV precision.

In this paper, our goal is to establish a Bayesian framework that safeguards the inference against confusion between baryonic feedback models. We aim to preserve accuracy in our cosmological constraints while retaining most of the cosmological information contained in our non-Gaussian estimator, fully leveraging Stage IV statistical power. In this paper, we focus on the peak counts of the convergence field. We account for the uncertainy in the baryonic feedback modeling by implementing Bayesian Model Averaging (BMA), a statistical framework that produces robust predictions for model parameters by combining each model's posterior distributions weighted by its relative probabilities of having generated the data.

Our paper is structured as follows. In Sect.~\ref{sec:data} we present our data analysis set up, including the mock weak lensing maps based on N-body simulations and the hydrodynamic simulations. We also present our strategy to account for the influence of baryons on non-Gaussian statistics. In Sect.~\ref{sec:bma} we introduce the formulation of Bayesian model averaging and how it can be applied to baryonic feedback models. In Sect.~\ref{sec:inference} we present our likelihood and the analysis set up at Stage IV precision. We finally show our results in Sect.~\ref{sec:results} followed by the conclusions in Sect.~\ref{sec:conclusion}.

\section{Data analysis setup}\label{sec:data}
In this paper, the aim is to infer primary cosmological parameters $\both$ from observed maps of weak gravitational lensing.  The weak lensing convergence fields $\kappa(\vartheta, \varphi)$ are mapped as a function of the celestial coordinates $\vartheta,\varphi$. They can be gained from observations of sheared galaxies with algorithms such as Almanac \citep{Sellentin:2023uou, Loureiro:2022fuc}, or competing mass-mapping algorithms \citep{1993ApJ...404..441K, Bartelmann:1995yq,1996ApJ...473...65S,2020A&A...638A.141P, 2022MNRAS.512...73F, 2024arXiv240305484B}.

Weak lensing convergence maps $\kappa(\vartheta, \varphi)$ are non-Gaussian random fields. Therefore, estimators beyond the  power spectrum are needed to maximise the information extracted from the non-Gaussianity contained in such maps. Technically, there are infinitely many non-Gaussianity estimators. In this paper, we study the peak counts of lensing convergence $\kappa$ maps \citep{Jain:1999nu, vanwaerbeke2000}. As shown in \cite{2011PhRvD..84d3529Y} and \cite{Liu:2016xjb}, the peak heights in convergence maps are direct tracers of massive haloes and projection of smaller haloes along the line of sight, which are sensitive to cosmological parameters. 
To measure the peaks, we count the number of pixels in convergence maps whose values are larger than the 8 neighboring pixels. This results in the distribution of local maxima in a convergence map, as a function of $\kappa$.

Our simulated convergence maps are based on SLICS \citep{Harnois-Deraps:2018bcv} and cosmo-SLICS N-body simulations \citep{2019A&A...631A.160H}. These 100 deg$^{2}$ maps mimic Stage IV  properties, such as shape noise and source redshift distribution. More details on the simulations and map production can be found in \cite{Daniela2023}. We consider five tomographic redshift bins in the ranges $0.25<z_1<0.75$, $0.75<z_2<1.25$, $1.25<z_3<1.75$, $1.75<z_4<2.25$ and $2.25<z_5<2.75$, with number densities $n_{\text{gal}}={18.27, 14.58, 7.89, 4, 1.89}$ arcmin$^{-2}$, respectively. Then, we include shape noise to our maps by adding to each pixel a value drawn from a Gaussian distribution centred at 0 with variance

\begin{equation}
\sigma_{\rm noise}=\frac{\sigma_{e}}{\sqrt{n_{\text{gal}}A_{\text{pix}}}},
\end{equation}
where $A_{\text{pix}}$ is the solid angle per pixel, and we adopt $\sigma_{e}=0.26$ for the mean intrinsic ellipticity. Finally, we apply a Gaussian kernel to smooth the maps with variances of 2 and 5 arcmin. Varying the smoothing scale allows us to suppress noise and exploit different features of the data. We choose 2 arcmin as this value is above the resolution of our maps while maintining the information of the small scales. Larger smoothing scales remove small structures, thereby reducing the effect of baryons to some extent.
Therefore, our analysis consists of multiple configurations, where peak counts are measured in 10 $\kappa$ bins, obtained for the five tomographic bins and smoothing scales.

\subsection{Baryon correction modelling}\label{sec:correction}
To date, there is no analytical expression to include baryonic effects in the modeling of the peak counts. Therefore, we instead introduce its effects into the dark matter-only estimators as a correction factor obtained from hydrodynamic simulations. We build convergence maps based on the BAHAMAS simulations \citep{Mccarthy:2017yqf, McCarthy:2016mry} at the WMAP nine-year cosmology \citep{2013ApJS..208...19H}. In particular, we include BAHAMAS runs with AGN heating temperature raised and lowered by 0.2dex with respect to the fiducial value. We refer to these models as ‘high-AGN’ (stronger feedback), ‘fid-AGN’ and ‘low-AGN’ (lower feedback). We also include the dark matter only counterpart, which we denote as ‘DMO’. For each of these models, we have 10,000 realizations obtained from 25 independent light cones, with 400 realizations each generated through random rotations and shifts of the potential planes.
To include the Stage IV properties, we follow the same methodology described for the SLICS and cosmo-SLICS convergence maps.

The correction factor $B$ is obtained for the three baryonic feedback scenarios. The elements $B_i$ of this factor are computed as follows
\begin{equation}
 B_i = \frac{\langle x_i^{\text{B}} \rangle}{\langle x_i^{\text{DMO}} \rangle},
 \label{baryon_factor}
\end{equation}
where the angular brackets represent an average over 10,000 realizations. The factor $\langle x_i^{\text{B}} \rangle$ denotes the average peak counts measured in hydrodynamic maps, while $\langle x_i^{\text{DMO}} \rangle$ denotes the average peak counts from the corresponding dark matter only maps. We opt for this approach instead of introducing estimators obtained from BAHAMAS mocks directly as any slight discrepancy between mocks cancels out to leading order when computing the ratio in Eq.~\ref{baryon_factor}. 

We show the impact of baryons on the peak counts for the five tomographic bins and smoothing scales in Appendix~\ref{app}. From Fig.~\ref{fig:ratios_peak}, we see that the high-AGN model produces the strongest effects on the peak counts, specially for the high $\kappa$ regime. This result is consistent with previous analyses presented in \cite{ Coulton:2019enn, Osato:2020sxo,2024MNRAS.529.2309B,Grandon:2024pek} based on convergence maps with different redshift bins, noise properties and hydrodynamic simulation. Studies of baryonic feedback on the peak counts based on the baryonic correction model also show the same overall effect \citep{yangpeak,Weiss:2019jfx}. Therefore, baryons can introduce biases in the cosmological parameters if their effects are not modelled correctly. This effect is less significant when increasing the smoothing scale, due to the removal of small-scale structures where baryons become important, and potentially a loss of precision in cosmological constraints.

To introduce the effect of baryons on the likelihood mean $\mu(\theta)$, we impose the product
\begin{equation}
 \mu_i^B = \mu_i(\both) B_i,
 \label{correction}
\end{equation}
 %\ref{sec:baryonic}
where we assume the fractional impact of baryonic feedback on the peak counts is cosmology-independent \citep{2024MNRAS.529.2309B}. We present more details on the correction factor and its implementation in the likelihood in Section~\ref{sec:inference}.

\section{Bayesian Model averaging and posterior setup}\label{sec:bma}

The standard practice in statistical inference for cosmology consists of assuming the existence of a cosmological model that could have generated the data, and then estimating the model parameters based on the observed data. The best-fit parameter values are thus conditioned on the chosen model. When multiple competing theoretical models exist, we can advance further and select the model that is favoured by the data according to some criteria, such as the evidence ratio in the Bayesian framework. Therefore, we draw conclusions assuming the selected model to be true. However, like many other prespecified models, this model may still be an approximation. This raises the question of how to address the fact that we select a model from a range of competing candidate models. A way to propagate this model uncertainty is to implement Bayesian model averaging (BMA). The BMA address the model uncertainty by performing an average of candidate models, producing more robust predictions (see  \cite{bma, 10.1214/ss/1009212519} for a review)\footnote{For previous applications of BMA in the study of cosmological models, we refer the reader to \cite{PhysRevD.74.123506, PhysRevD.82.103533,Vardanyan:2011in,Paradiso:2024pcb, Paradiso:2023ohr}.}. In the average each model posterior probability is weighted by its Bayesian evidence.

The BMA posterior corresponds to
\begin{equation}
\calP( \boldsymbol{\theta} | \fatx ) = \frac{\sum_k \calP(\boldsymbol{\theta}|\fatx,M_k)\calP(M_k|\fatx) }{\sum_k \calP(M_k | \fatx)}
\label{eqbma}
\end{equation}
where $\calP(M_k|\fatx)$ is the Bayesian evidence of model $M_k$ and $\calP(\boldsymbol{\theta}|\fatx,M_k)$ is the posterior of each model $M_k$. The evidences are given by
\begin{equation}
    \calP(M_k|\fatx) = \int \calL(\fatx|\both_k,M_k) \pi(\both_k|M_k) \mathd^{r_k} \theta_k,
    \label{evi}
\end{equation}
where $\both_k$ are all $r_k$ parameters that model $M_k$ uses. As model $M_k$ might use more parameters than the primary cosmological parameters, $r_k$ can differ per model. In order to arrive at a model averaged posterior for parameters $\both$, the vectors $\both_k$ however include all parameters of $\both$. The term $\calL$ in Eq.~(\ref{evi}) is the likelihood, and $\pi$ is the prior on the model parameters $\both_k$ in model $M_k$.

The evidence expresses the total probability that a model $M_k$ has generated the data $\fatx$ \emph{at all}. As can be seen, the evidence integrates over all parameters of model $M_k$ and weighs their contribution to the total evidence by the likelihood and the prior probability. If model $M_1$ has a larger evidence than model $M_2$ in light of the data $\fatx$, then model $M_1$ is more likely to be the model to have generated the data. By evaluating evidences, one can therefore rank models by their relative probability to have generated the data.

The complexity of baryonic physics in the large-scale structure has led to multiples approaches to model its effects. Therefore, the BMA is a natural solution to address this model uncertainty in baryonic physics for weak lensing inference. We implement the BMA in the analysis of non-Gaussianity estimators so that the data can determine which baryonic feedback model is the most likely.

Finally, we compute the Bayes factor between models. Given two competing models $M_j$ and $M_k$, the ratio of the evidences corresponds to \citep{bayesfactors, jeffreys1998theory}
\begin{equation}
BF_{jk}=\frac{\calP(M_j|\fatx) }{\calP(M_k|\fatx) },
\end{equation}
with models having the same prior probability.

\begin{figure*}
\centering

   \begin{tabular}{ccr}
              \centering
            \includegraphics[width=\columnwidth]{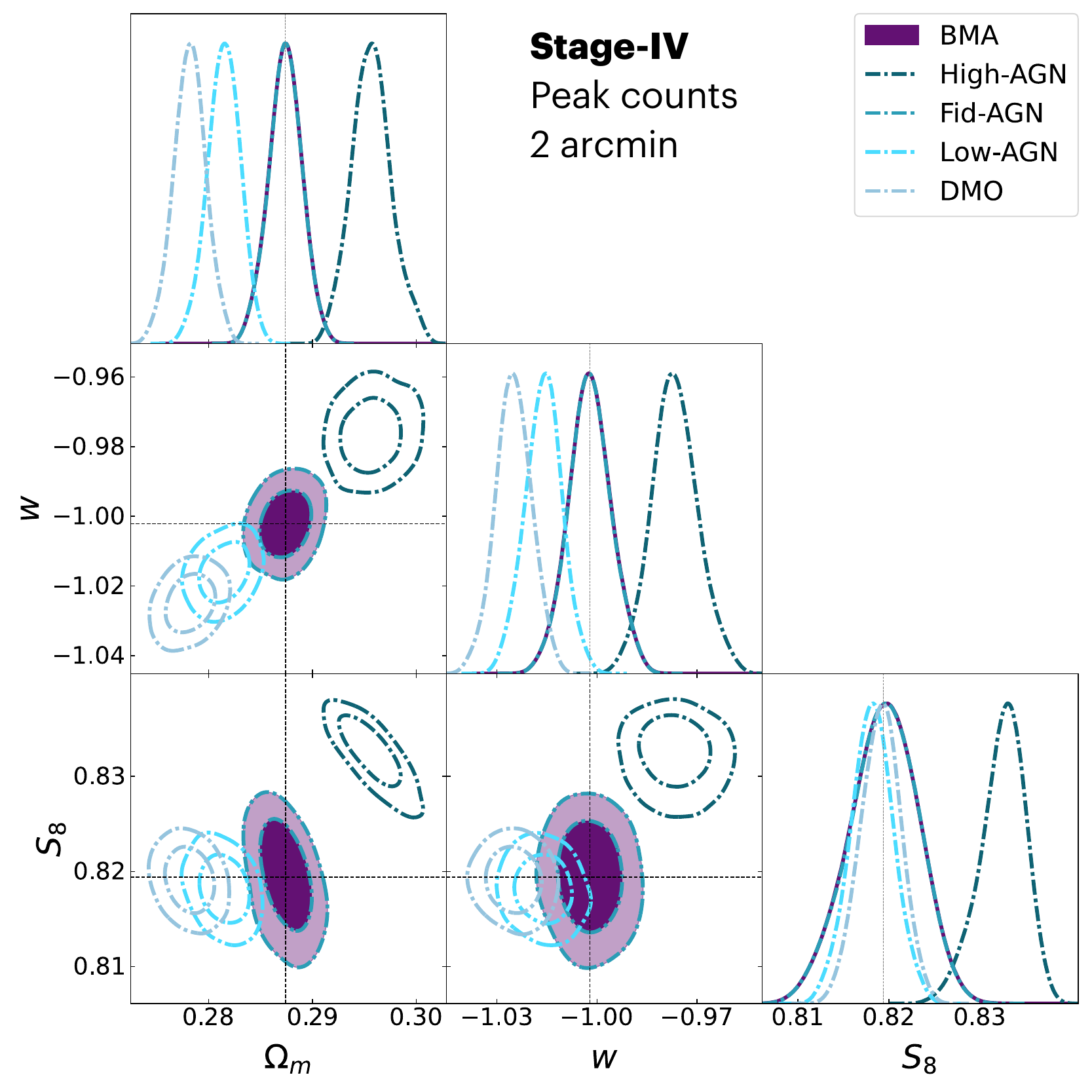}
        \hfill
            \includegraphics[width=\columnwidth]{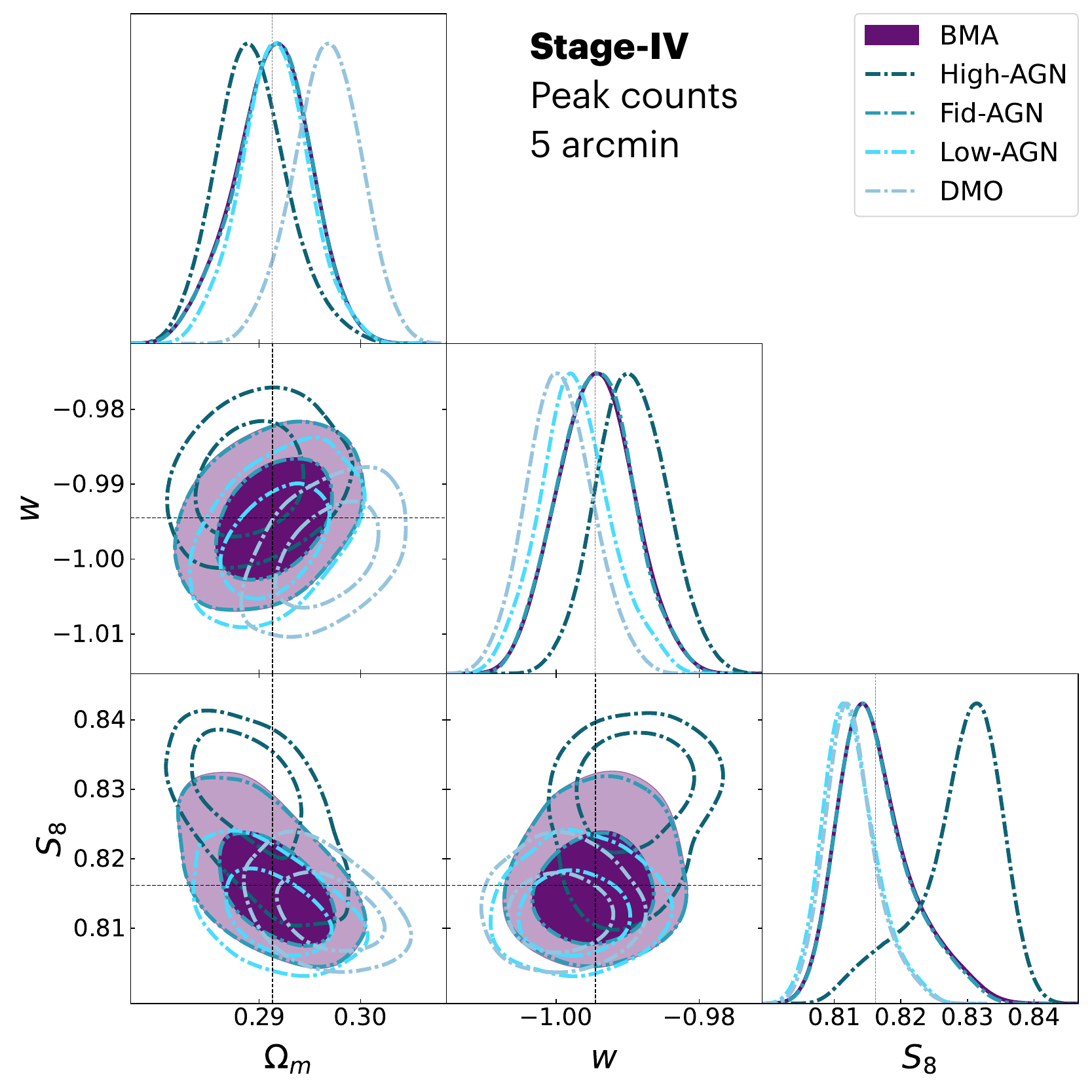}
        \end{tabular}
  
            \caption 
        {Cosmological constraints of $S_8$, $w$, and $\Omega_m$ based on the peak counts with `Case 1' fid-AGN data vector. \textit{Left:}  Results obtained for convergence maps smoothed with a Gaussian kernel of 2 arcmin smoothing scale. \textit{Right}: Results with a Gaussian kernel of 5 arcmin smoothing scale. Dashed line denotes the input true cosmology. The three baryonic feedback models and dark matter only model are depicted with dashed contour lines. The Bayesian model averaging result, shown in purple with solid contour lines, combines all baryonic feedback models and successfully recovers the true cosmological parameters.} 
        \label{fig:results_2arcmin}
\end{figure*}

\begin{figure*}
\centering

   \begin{tabular}{ccr}
              \centering
            \includegraphics[width=\columnwidth]{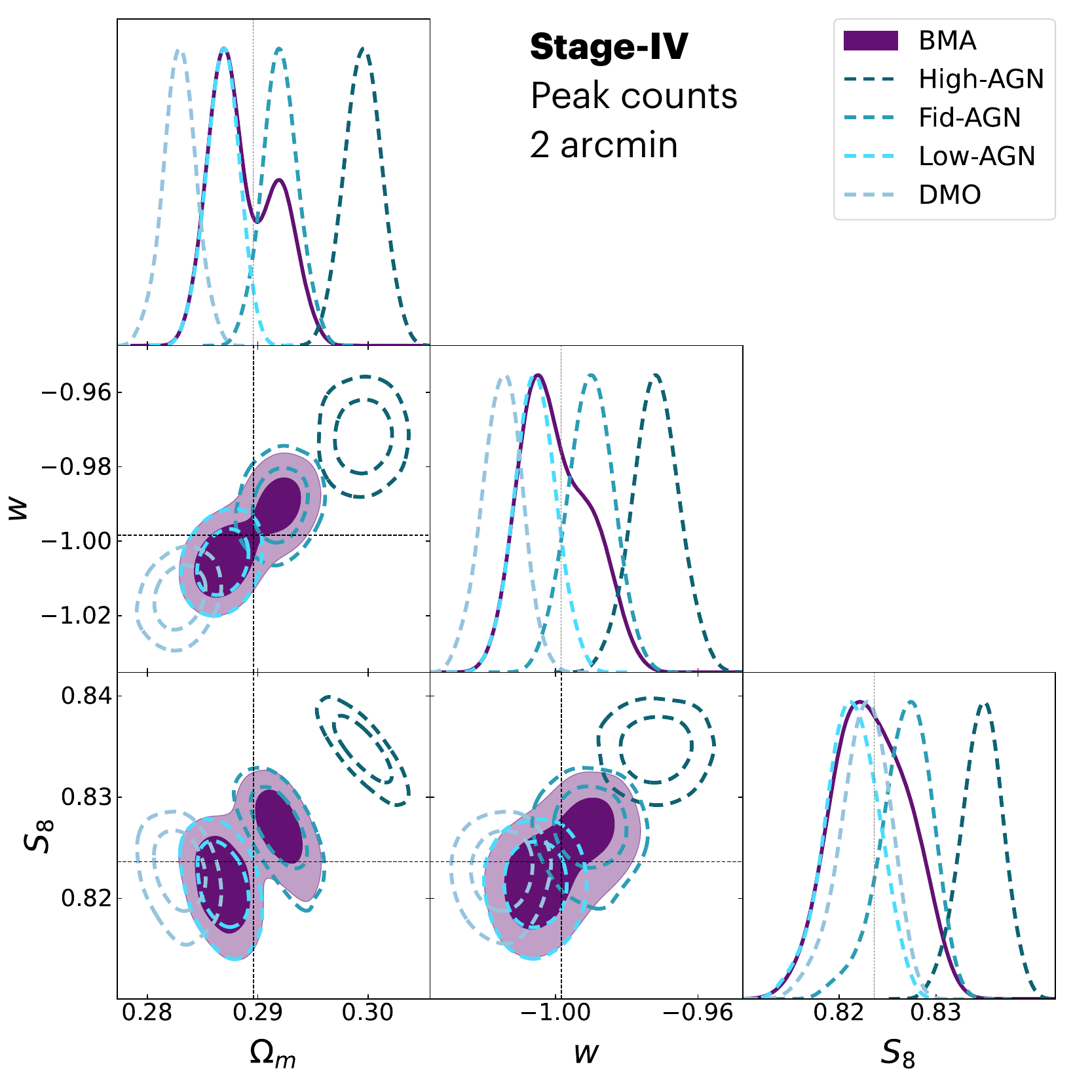}
        \hfill
            \includegraphics[width=\columnwidth]{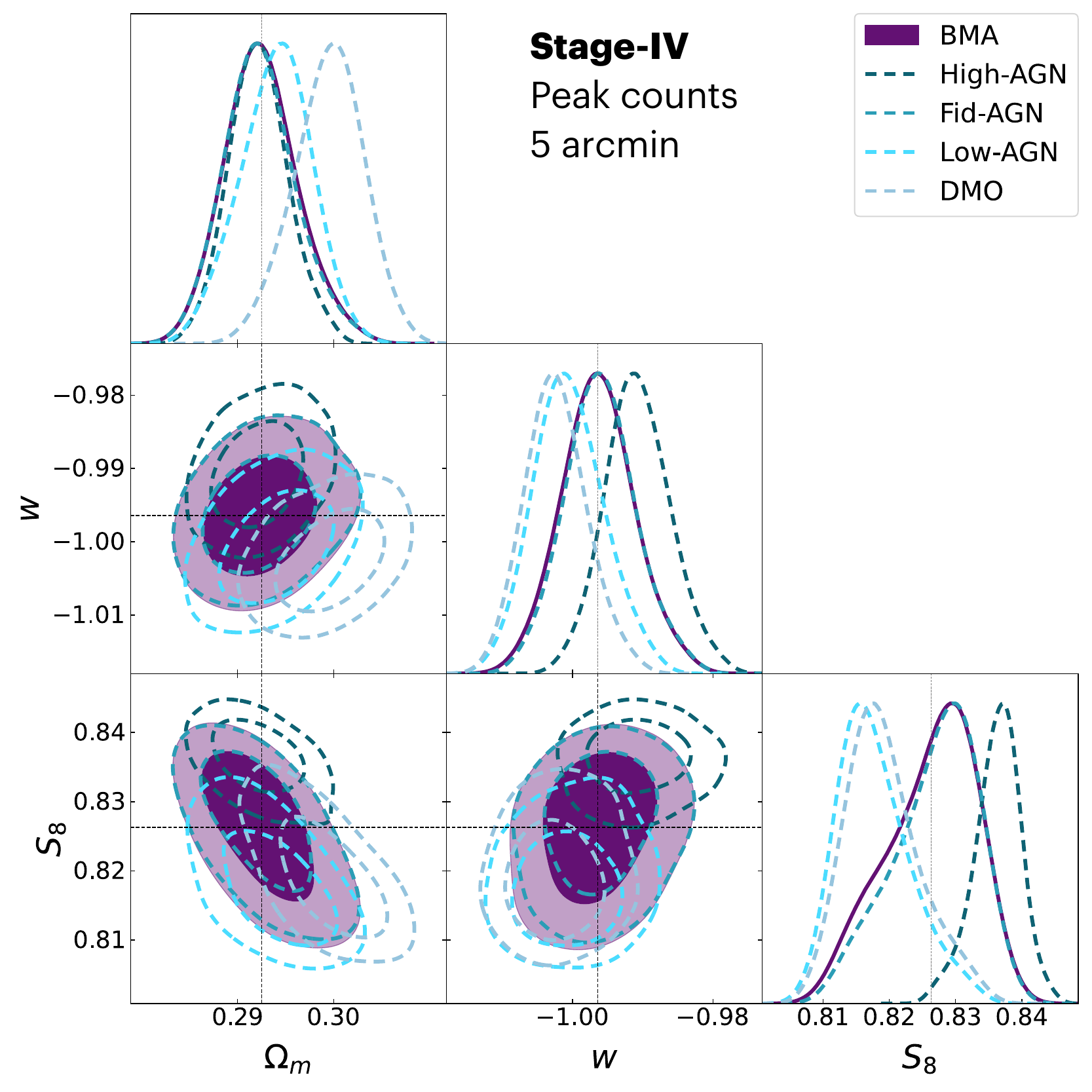}
        \end{tabular}
  
            \caption 
        {\small Same as Fig.~\ref{fig:results_2arcmin} for the model misspecification data vector. In this case, none of the baryonic feedback models provides an accurate approximation of the influence on baryons as presented in the data vector. This results in the bias observed for all models. However, the BMA result (shown in purple contour) successfully recovers the true cosmological parameters by combining the four posteriors.} 
        \label{fig:results_5arcmin}
\end{figure*}

\section{Parameter inference}\label{sec:inference}
This section describes the setup of our posterior.
We first estimate the covariance matrix from the SLICS simulations. It is computed as follows
\begin{equation}
    \boC = \frac{1}{N_r -1} \sum_{n=1}^{N_{r}} (\boX_{n} - \bar{\boX})(\boX_{n} - \bar{\boX})\,,
\end{equation}
where $N_r=953$ is the number of realizations per estimator $\boX$ at the fiducial cosmology, and $\bar{\boX}$ the mean for the estimators. We assume a Stage IV survey with a sky coverage of 18,000 deg$^{2}$, and thus we rescale the covariance matrix as $\boC = (100/18,000)\boC $. For estimated covariance matrices we adopt the likelihood function presented in \cite{Sellentin:2015waz}. This corresponds to the modified t-distribution

\begin{equation}
P(\fatx_o | \bomu_B, \boC, N_r) \propto \left[ 1 + \frac{(\fatx_o -\bomu_B)^T\boC^{-1} (\fatx_o - \bomu_B) }{N_r-1}\right] ^{\frac{-N_r}{2}},
  \label{cosmo_tdistrib}
\end{equation}
where $\mu_B$ corresponds to the mean including the effect of baryons in eq.~\ref{correction}. To obtain $\mu(\both)$, we model the peak counts for arbitrary cosmologies by training a Gaussian Process emulator, implemented in \textit{scikit-learn}\footnote{\url{https://scikit-learn.org}} \citep{scikit-learn}. Our training set consist of the 26 cosmo-SLICS cosmologies in the parameter space of $\Omega_m$, $w$ and $S_8$, for which we calculate the peak counts. We implement a Radial Basis Function kernel and check its accuracy using a leave-one-out cross-validation test. The emulator errors are below $1\sigma$ uncertainty of the survey, however the training of emulators is still an open challenge for Stage IV precision. 
We further describe the emulator challenges for various non-Gaussian statistics and the effects of the error propagation (e.g. \cite{Grandón2022Bayesian,Harnois-Deraps2024}) in \cite{Daniela2023}. 
Thorough this paper, we implement flat prior probability for the parameters given by $-1.80 < w < -0.70$, $0.63 < S_8 < 0.89$ and $0.1 < \Omega_m < 0.55$.

Our estimators are obtained from gravity-only simulations. Hence, to include baryonic effects in the theory $\bomu_B$, we infuse the effect of baryons into the dark matter only mean, as presented in Eq.~\eqref{baryon_factor}.

\subsection{Data vector set up}

To emulate a real science case at the Stage-IV precision, we infuse the baryons on the data vector as well. First, we assume this data vector is drawn from a Gaussian $\calG$ as
\begin{equation}
    \fatx \sim \calG(\bomu(\both_t),\sfC),
\end{equation}
where $\sfC$ is the data covariance matrix and the data's expectation value equals $\langle \fatx \rangle = \bomu(\both_t)$, i.e. a parametric mean evaluated at the position of the true cosmological parameters $\both_t$. As real data contains baryonic feedback, we correct for the baryons into the data vectors following the same procedure as for the mean. 

We explicitly write out the baryonic feedback model $M$ from which these data arise. $M$ can be `high AGN', `fid AGN',`low AGN', or `DMO'. Our data $\fatx$ and mean $\bomu$ always contain the same non-Gaussianity estimators, and the mean is always evaluated for each of the four models.
Obviously, if a data vector $\fatx$ stems in reality from model $M = \text{`high\ AGN'}$, but is then fitted with a mean $\bomu(\both)$ from model $M = \text{`low\ AGN'}$, then the inferred parameters $\both$ will be biased. This bias ensues from the incorrect baryon model being chosen. This is a significant concern of actual data analysis where the true impact of baryons on the large scale structure is far from sufficiently understood. 
To include baryonic physics into the data vector, we consider two cases for $\fatx$: 1) a data vector with fiducial AGN-like baryons; 2) and a data vector with model misspecification. The details on how to generate such cases are detailed below.

\subsubsection{Case 1: Data vector with fiducial AGN}
Our first case considers parameter inference with the data vector corresponding to $\obsboX= \fatx_iB_i$ where $B_i$ is the correction factor derived from the fiducial-AGN model. We repeat our parameter inference with this data vector, but with the corrected mean varying the model $M$.

\subsubsection{Case 2: Model misspecification}
If one proposes multiple models for fitting the data, then it may happen that none of them is the model that generated the data. This situation is called model misspecification. We imitate this situation by generating a data vector $\fatx_o$ as
\begin{equation}
    \fatx_o = (1-\lambda)\fatx(\text{fid-AGN}) + \lambda \fatx(\text{low-AGN})\,,
\end{equation}
with $\lambda=0.5$.
Hence, the correct mean of this data vector is 
\begin{equation}
    \bomu_B = (1-\lambda)\bomu(\text{fid-AGN}) + \lambda \bomu(\text{low-AGN}).
\end{equation}
To demonstrate model misspecification we purposefully fit the data with the four means of our original four models\footnote{We refer the reader to \cite{Porqueres:2023drp} for model misspecification in intrinsic aligment studies.}. 

We sample the posterior of the cosmological parameters $\Omega_m$, $S_8$, and $w$ with $\textsc{MultiNest}$ \citep{Feroz:2008xx} as implemented by $\textsc{PyMultiNest}$ \citep{2014A&A...564A.125B}. $\textsc{MultiNest}$ reports the Bayesian evidence alongside the parameter constraints. We run our analysis for the two data vectors cases considered and two smoothing scales. To study the model preferred by the data, we calculate the logarithm of the Bayes factor.

\section{Results}\label{sec:results}

\subsection{Case 1}

We report our results and logarithm of the Bayes factors in Table~\ref{bayesratio1} and Fig.~\ref{fig:results_2arcmin}. Here, the data vector stem from the fiducial-AGN model (case 1) for 2 and 5 arcmin smoothing scales. 
In Fig.~\ref{fig:results_2arcmin}, we demonstrate how to remove the bias from the cosmological inference. 
The four posteriors in open contours depict the posteriors of each individual model, where three of them are biased posteriors respect to the true cosmology (in dashed black lines). The biases arise from fitting three incorrect models to this data (DMO, high AGN and low AGN), followed by the true baryonic feedback model centered at the true cosmology. The biases in the inferred cosmological parameters are statistically significant and reach  $\sim-4\sigma$ for $\Omega_m$ and $\sim3\sigma$ for $w$ and $S_8$ when the modeling in the mean is incorrect. 
From the Bayes ratio in Table~\ref{bayesratio1}, we see the data have discriminating power between these different baryonic models. The logarithm of the Bayes ratio ranges from 6 up to 12.35 when compared to the fid-AGN model. According to Jeffreys' scale, this indicates decisive evidence in favor of this baryonic feedback model. We therefore evaluated BMA posterior from Eq.~\ref{eqbma} and display it in filled purple contours. The model-averaged (the BMA) posterior is centred at the true cosmology, and thus removes biases in the cosmological constraints. In this particular case, it is also almost identical to the posterior of the correct fid-AGN baryon model. This shows that for strongly constraining data, the analysis succeeds in identifying the preferred model. The other competing models are then downweighed due to their inferior evidence, as seen in Table~\ref{bayesratio1}.\\

From Table~\ref{bayesratio1} for 5 arcmin, we see that there is also decisive evidence in favor of the fid-AGN model when it is compared to the other models. The 5 arcmin posterior results correspond to the right corner plot in Fig.~\ref{fig:results_2arcmin}. We observe that all baryonic feedback models are less biased compared to the 2 arcmin smoothing scale. This is because, as we increase the smoothing scale, we lose precision and we partially remove the imprints of baryons at the smallest scales of the maps. This is supported by the ratio of the peak counts with and without baryonic physics in Fig.~\ref{fig:ratios_peak}, where the impact of baryons is less significant for the last three tomographic bins. This, in turn, makes the evidence of the individual feedback models closer to each other. This results in a wider BMA posterior contour (contour in filled purple), as seen in Fig.~\ref{fig:results_2arcmin}. We can therefore obtain accurate cosmological results, though with some loss of precision due to the propagation of baryonic feedback model uncertainty. Still, this allows us to make statistically robust claims about the inferred parameters.

\subsection{Case 2}
Fig.~\ref{fig:results_5arcmin} shows our results for the model misspecification case, where the baryons in the data vector do not correspond to any of the baryonic feedback models. In this case, the data then have less preference for a single model, and instead the evidences of the two closest models dominate. This is indicated by the results in Table~\ref{bayesratio2}, both for 2 and 5 arcmin smoothing scales. The posterior gained from Bayesian model averaging hence combines the two closest posteriors, as seen in Fig.~\ref{fig:results_5arcmin}. Multi-modality in the posterior is an expected feature in Bayesian model averaging, and we here observe it in the resulting posterior for $\Omega_m$ for 2 arcmin result. As can be seen, although the true cosmology was excluded by the four biased posteriors, it is correctly assigned posterior credibility by the model-averaged posterior. This multimodality is not seen in the 5 arcmin smoothing scale, as all contours are less biased respect to the true cosmology. For 2 arcmin, the marginal distributions of $S_8$ and $w$ for the BMA exhibit a higher concentration of probability mass in the right tail of the distribution, due to the influence of the fid-AGN model in the model average.

\begin{table*}
\centering
\begin{tabular}{|ccccc|}
\hline
   %                        &                             & \multicolumn{1}{l}{\textbf{Scale cuts/Smoothing scales}}       &                            \\
\multicolumn{1}{|l}{} \rule{0pt}{2.2ex}  & DM            & BAHAMAS low-AGN                               & BAHAMAS fid-AGN  & BAHAMAS high-AGN  \\[0.3ex] 
 \hline
 \rule{0pt}{2.1ex}   & &  2 arcmin& &\\ [0.2ex] \hline
 
\rule{0pt}{2.05ex}  DM & 0                     & -6.23 & -12.35 & -1.99         \\ \rule{0pt}{2.1ex}  
BAHAMAS low-AGN           &    6.23      & 0           & -6.11      & 4.23         \\\
\rule{0pt}{2.1ex}   BAHAMAS fid-AGN              &  12.35 &  6.11   & 0     & 10.35    \\ \rule{0pt}{2.1ex}  

BAHAMAS high-AGN    &  1.99     & -4.23 & -10.35           & 0  
\\[0.2ex] 
 \hline
 \rule{0pt}{2.1ex}   & &  5 arcmin& &\\ [0.2ex] \hline
 \rule{0pt}{2.1ex} DM & 0  & -0.71 & -2.91 & -0.93   \\ 
 \rule{0pt}{2.1ex}  
BAHAMAS low-AGN   &  0.71  & 0           & -2.21  & -0.22   \\\
\rule{0pt}{2.1ex}   BAHAMAS fid-AGN   &  2.91 &  2.21   & 0     & 1.98    \\ \rule{0pt}{2.1ex}  

BAHAMAS high-AGN    &  0.93     & 0.22 & -1.98           & 0   
\rule[-1.1ex]{1pt}{0pt}   \\

\hline

\end{tabular}
  \caption{Logarithm of Bayes factor for the baryonic feedback models considered in this work. This table shows the results obtained from the fiducial-AGN data vector (case 1).} 
\label{bayesratio1}
\end{table*}

\begin{table*}
\centering
\begin{tabular}{|ccccc|}
\hline
   %                        &                             & \multicolumn{1}{l}{\textbf{Scale cuts/Smoothing scales}}       &                            \\
\multicolumn{1}{|l}{} \rule{0pt}{2.1ex}  & DM            & BAHAMAS low-AGN                               & BAHAMAS fid-AGN  & BAHAMAS high-AGN  \\[0.2ex] 
 \hline
 \rule{0pt}{2.1ex}   & &  2 arcmin& &\\ [0.4ex] \hline
\rule{0pt}{2.1ex}  DM & 0  & -4.77 & -4.54 & 10.90   \\ \rule{0pt}{2.1ex}  
 \rule{0pt}{2.1ex}  
BAHAMAS low-AGN   &  4.77  & 0           & 0.23  & 15.68   \\\
\rule{0pt}{2.1ex}   BAHAMAS fid-AGN   &  4.54 &  -0.23   & 0     & 15.45    \\ \rule{0pt}{2.1ex}  

BAHAMAS high-AGN    &  -10.90     & -15.68 & -15.45           & 0  
\\[0.2ex] 
 \hline
 \rule{0pt}{2.1ex}   & &  5 arcmin& &\\ [0.2ex] \hline
 \rule{0pt}{2.1ex} DM & 0  & -1.56 & -2.69 & 0.73   \\ \rule{0pt}{2.1ex}  
 \rule{0pt}{2.1ex}  
BAHAMAS low-AGN   &  1.56  & 0           & -1.13  & 2.29   \\\
\rule{0pt}{2.1ex}   BAHAMAS fid-AGN   &  2.69 &  1.13   & 0     & 3.42    \\ \rule{0pt}{2.1ex}  

BAHAMAS high-AGN    &  -0.73     & -2.29 & -3.42           & 0   
\rule[-1.1ex]{2pt}{0pt}   \\

\hline

\end{tabular}
  \caption{Same as Table \ref{bayesratio1}, but for the case 2 of model misspecification data vector.} 
\label{bayesratio2}
\end{table*}

\section{Conclusions}\label{sec:conclusion}

This paper presents a Bayesian solution to safeguard parameter inference against biases resulting from baryonic feedback not being correctly modeled in the weak lensing non-Gaussian statistics. Our solution consists of Bayesian model averaging, a statistical framework that proposes a posterior distribution combining the individual posteriors of multiple competing models that could have generated the observed data. Therefore, instead of comparing models by means of model selection criteria, the BMA enables more robust predictions by averaging all models, and hence propagating this model uncertainty. The BMA posterior is presented in Eq.~\ref{eqbma}. 

In this paper we focus on three baryonic feedback models that impact the non-Gaussian estimators and hence the inference of cosmological parameters $\Omega_m$, $w$ and $S_8$. We perform a tomographic analysis of the convergence peak counts at Stage IV precision.

Our results are shown in the Figures~\ref{fig:results_2arcmin} and \ref{fig:results_5arcmin}. Fig.~\ref{fig:results_2arcmin} corresponds to the resulting posteriors when the data vector is corrected by the fiducial-AGN model, which also corresponds to one of the theory models. Fig.~\ref{fig:results_5arcmin} corresponds to our results when the data vector is none of the baryonic feedback models, and hence represent a model misspecification case. As can be seen from the Bayes factors in Tab.~\ref{bayesratio1} and~\ref{bayesratio2}, the cosmological data at Stage IV precision can distinguish between different feedback models, making them more or less a good fit. This means the data will suppress bad models; and hence the largest evidence is for the correct model. By averaging the posteriors obtained from all baryonic feedback models, weighted by their evidence, the resulting BMA posterior correctly finds the true cosmology within the 68\% C.L. 
This demonstrates that fitting a multitude of baryon models to the data, and having the data select the best models, is a solid technique to accomplish accuracy in the inference from non-Gaussianity estimators. 

\section*{Acknowledgements}

We thank Ian McCarthy and the BAHAMAS simulation team for making their simulations publicly available. We would like to thank Andrew Jaffe, Alan Heavens, Joachim Harnois-Déraps, Jia liu, Javier Silva-Lafaurie, Kutay Nazli and Tatiana M. Rodriguez for useful discussions.

\section*{Data AVAILABILITY}

The data underlying this article will be shared on reasonable request to the corresponding author.

\bibliographystyle{mnras}
\bibliography{peakcounts}

\appendix
\section{Appendix A}\label{app}
\begin{figure*}
                        \centering
            \includegraphics[width=\textwidth]{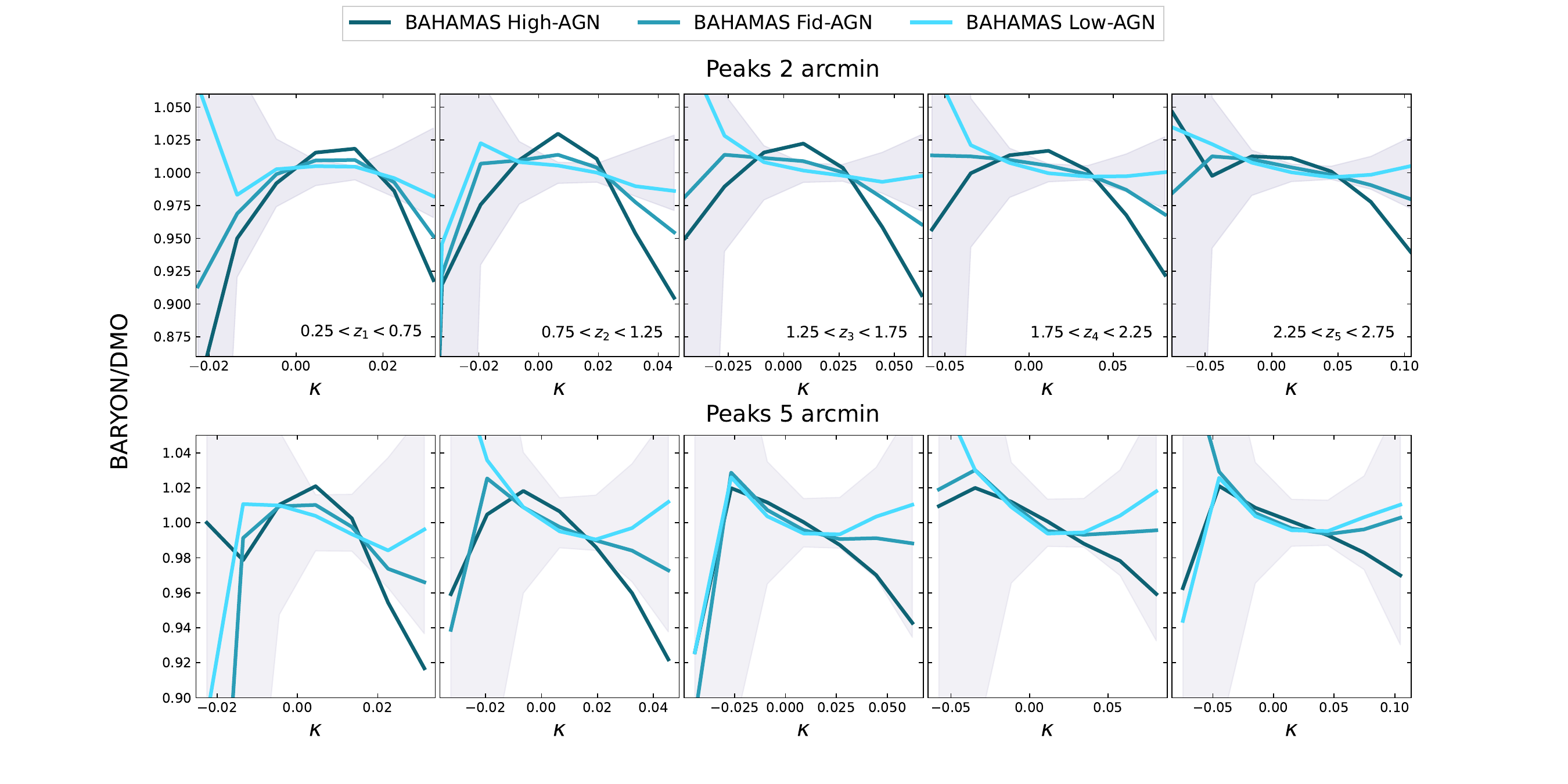}

            \caption 
        {\small The impact of baryonic feedback on the peak counts for smoothing scales of 2 arcmin (top) and 5 arcmin (bottom). The panels from left to right show the results for the tomographic bins. The grey shaded region indicates the survey $1\sigma$ uncertainty.} 
        \label{fig:ratios_peak}
\end{figure*}

The impact of baryons on the peak counts is presented in Fig.~\ref{fig:ratios_peak}. The grey shaded region corresponds to Stage IV 1$\sigma$ uncertainty, obtained from the diagonal of the data covariance matrix in Eq. 7. We find that baryonic feedback reduces the number of peaks up to 10\% for large $\kappa$, and two out of three of our baryonic feedback models produce effects that exceeds the error budget of the survey. Therefore, neglecting the effects of baryons can lead to statistically significant biases in cosmological constraints, as confirmed by our results presented in Figures \ref{fig:results_2arcmin} and \ref{fig:results_5arcmin}.

% Don't change these lines
\bsp	% typesetting comment
\label{lastpage}
\end{document}